# SENSITIVITY OF PDR CALCULATIONS TO MICROPHYSICAL DETAILS

N. P. Abel[1], P. A. M. van Hoof[2], G. Shaw[3], G. J. Ferland[4] & T. Elwert[4]

[1]Department of Physics, University of Cincinnati, Cincinnati, OH, 45221 npabel2@gmail.com

[2]Royal Observatory of Belgium, Ringlaan 3, 1180 Brussels, Belgium; p.vanhoof@oma.be

[3]Department of Astronomy and Astrophysics, Tata Institute of Fundamental Research, Mumbai 400 005, India; gargishaw@gmail.com

[4]University of Kentucky, Department of Physics and Astronomy, Lexington, KY 40506 gary@pa.uky.edu; elwert@avrte.com

## Abstract

Our understanding of physical processes in Photodissociation regions or Photon Dominated Regions (PDRs) largely depends on the ability of spectral synthesis codes to reproduce the observed infrared emission-line spectrum. In this paper, we explore the sensitivity of a single PDR model to microphysical details. Our calculations use the Cloudy spectral synthesis code, recently modified to include a wealth of PDR physical processes. We show how the chemical/thermal structure of a PDR, along with the calculated spectrum, changes when the treatment of physical processes such as grain physics and atomic/molecular rates are varied. We find a significant variation in the intensities of PDR emission lines, depending on different treatments of the grain physics. We also show how different combinations of the cosmic-ray ionization rate, inclusion of grain-atom/ion charge transfer, and the grain size distribution can lead to very similar results for the chemical structure. Additionally, our results show the utility of Cloudy for the spectral modeling of molecular environments.

## 1 Introduction

Photodissociation regions or Photon Dominated Regions (PDRs) are regions where hydrogen makes the transition from H to $H_2$, and where other molecules such as CO form. PDRs contain the majority of mass in regions of star-formation, making them ideal laboratories to study physical processes in astrophysical environments (Hollenbach & Tielens 1997). PDRs are often physically adjacent to H II (or $H^+$) regions, although environments with little hydrogen-ionizing radiation, such as reflection nebulae, also contain PDRs (Hollenbach & Tielens 1997; Young Owl et al. 2002).

Determining the physical conditions in PDRs usually involves combining spectroscopic observations with theoretical calculations (Tielens & Hollenbach 1985; Wolfire, Tielens, & Hollenbach 1990; Kaufman et al. 1999). PDRs are most conveniently studied at infrared, sub-mm, or radio wavelengths, where the effects of extinction are usually minimal and where atomic, molecular, and dust emission features are readily observable. Theoretical calculations involve solving the problem of radiative transfer for a cloud of gas and dust exposed to UV radiation. Two parameters typically define the calculation; the far ultraviolet (FUV) radiation field (parameterized by $G_0$, where $1G_0$ equals a flux of $1.6 \times 10^{-3}$ erg cm$^{-2}$ s$^{-1}$, integrated over the energy range 6 – 13.6 eV, Habing 1968) and total hydrogen density $n_H$ [where $n_H \approx n(H^0) + 2n(H_2)$]. Included in the computational methods are treatments of all the various physical processes which the predicted spectrum sensitively depends upon. Some of the more important physical processes include grain physics, molecule formation, photoionization/photodissociation, and cosmic ray/x-ray heating, and ionization. This is in addition to model assumptions about the geometry and abundances.

In this paper, we show how the results of a single PDR calculation (using fixed values for $n_H$ and $G_0$) vary as various physical processes are included or excluded. In Section 2, we describe important physical processes which must be treated in every PDR calculation. In Section 3, we describe our computational details. We present our results in Section 4, and give a summary of our conclusions in Section 5.

## 2 Physical Processes in a PDR

### 2.1 Dust Physics

Dust physics in a PDR is arguably the most important physical process controlling the chemical/thermal structure. Photoelectric heating of dust is often the primary heating agent in a PDR (Bakes & Tielens 1994; Weingartner & Draine 2001a), with single photon heating also contributing to dust heating (Sellgren 1984; Guhathakurta & Draine 1989). If the density is sufficiently high, collisions between the gas and dust at different temperatures can heat the gas (see Hollenbach & McKee 1979; Tielens & Hollenbach 1985). Other possible ways grains can heat the gas include viscous heating (Tielens & Hollenbach 1985), and absorption of the dust-emitted Far-Infrared (FIR) continuum by optically thick FIR emission lines, which are then collisionally de-excited (Tielens & Hollenbach 1985). Dust can also cool the gas through collisions, if the gas temperature ($T_g$) exceeds the dust temperature ($T_d$). Dust absorbs and scatters UV photons, making it an important opacity source in PDRs. The opacity effects of dust significantly affect photoionization/dissociation rates, thereby altering the chemical/ionization structure in a PDR. Dust affects the chemistry through the formation of $H_2$, which predominately forms on grain surfaces with a rate that depends on $T_g$, $T_d$, and the grain size (see Cazaux & Tielens 2002). The chemistry also depends on the dust through heating/cooling, since $T_g$ in the PDR is largely controlled by dust and many chemical rates depend on $T_g$. In cold ($T_g$, $T_d$ < 20 K) environments, molecules such as CO and $H_2O$ freeze out of the gas phase onto grain surfaces (see Bergin et al. 1995), a process which depends on $T_g$, $T_d$, and the dust grain size. Finally, charge exchange between gas and dust plays an important role in the ionization balance (Draine & Sutin 1987; Lepp et al. 1988; Weingartner & Draine 2001c). This can change the free electron density in a PDR as well as ionic abundances, which affects the chemical structure through dissociative recombination reactions with molecular ions.

## 2.2 Gas Phase Chemical Reactions

### 2.2.1 *Reactions Between Two Species*

Molecular abundances in a PDR are very sensitive to molecular reaction rates. Reaction rates depend on the quantum mechanics governing how reactants interact along with the local conditions of the gas. A detailed treatment of the physics of both is therefore necessary to determine the physical conditions in a PDR. Solving for molecular abundances is further complicated by the fact that molecular formation/destruction and the gas temperature/radiation field are coupled problems.

The most common type of reaction is one where both reactants are an atom/ion, molecule, or electron. The equation governing the rate coefficient is usually expressed as (see Prasad & Huntress 1980):

$$R = a\left(\frac{T_g}{300}\right)^b e^{\left(\frac{-c}{T}\right)} \text{ (cm}^3\text{ s}^{-1}) \tag{1}$$

where *a*, *b*, and *c* are coefficients which depend upon the interaction of the two reactants. Since most reactions involving two neutrals have a strong temperature dependence, and because the temperature in a PDR is of the order of a few hundred K, the fastest reactions in a PDR are usually electronic recombination and ion-neutral reactions (see Hollenbach & Tielens 1997 and references within). Such reactions, due to their electrostatic potential, typically do not have an exponential temperature barrier (*c* = 0).

### 2.2.2 *Photorates of Atoms and Molecules*

For photodissociation or photoionization reactions, the formally correct solution is to integrate the cross section for the photo process over the appropriate energy range times the local continuum intensity. To increase computational speed (particularly for photodissociation rates) it is often convenient to express the photorate $R$ as (see Roberge et al. 1991):

$$R = \alpha G_0 e^{(-\beta A_V)} \text{ (s}^{-1}\text{)} \tag{2}$$

where the parameters $\alpha$ and $\beta$ are parameters that depend on the cloud thickness and photodestruction process, and $A_V$ is the visual extinction. The exponential term accounts for the average attenuation of the $G_0$ continuum by dust.

Approximating the photorates using equation 2, while generally successful in reproducing observations, must be modified in some circumstances. Since the coefficient $\beta$ depends strongly on dust properties (van Dishoeck 1988), which varies depending on the environment, care must be taken to use the value of $\beta$ consistent with the region of interest. Equation 2 also does not take self-shielding of the UV radiation field by the molecule into account. Therefore, for the two most abundant molecules in space, $H_2$ and CO, equation 2 is always modified by various "self-shielding functions" that take into account various opacity effects of $H_2$ that shield $H_2$ and CO, such as those mentioned by van Dishoeck & Black (1988). These functions take into account the fact that, once the dissociating lines of $H_2$ become optically thick, the molecule shields itself against further photodissociation (Draine & Bertoldi 1996, van Dishoeck & Black 1987, Shaw et al. 2005). Therefore, for $H_2$ and CO, equation 2 is usually scaled by an extra factor (Draine & Bertoldi 1996). However, in some environments, this approximation has trouble reproducing observations; in particular diffuse atomic and translucent clouds that are illuminated by strong UV radiation fields (Abel et al. 2004, 2006; Le Petit et al. 2006, Snow & McCall 2006). These factors point to a need to either make sure the various approximations to the photorates are appropriate for the particular region of interest, or to use a detailed treatment of the $H_2$ self-shielding process, such as in van Dishoeck & Black (1987), Sternberg & Dalgarno (1989), Bertoldi & Draine (1996), Shaw et al. (2005), and Le Petit et al. (2006).

### 2.2.3 *Cosmic Rays*

In deep regions of a PDR, which are well shielded from UV radiation, the only significant source of ionization is high-energy cosmic rays or x-rays. Cosmic rays will directly ionize an atom or molecule through collisions (Dalgarno et al. 1999). Additionally, the electrons ejected through direct cosmic ray collisions are energetic enough to produce secondary ionizations (Dalgarno et al. 1999). Both processes are an important heating agent deep in PDRs. Cosmic rays can also excite atoms/molecules; most importantly H and $H_2$, which upon de-excitation will produce a small UV radiation field which can dissociate/ionize the surrounding environment (Gredel et al. 1989).

Uncertainties in the cosmic-ray ionization rate ($\xi$, which we define as the cosmic ray ionization rate of $H_2$) will affect the predicted abundances of molecules. Since cosmic rays are an important source of electrons in dense molecular gas, and dissociative recombination reactions are fast, the cosmic ray ionization rate will significantly affect the chemistry. For example, in diffuse clouds, the $H_3^+$ abundance is thought to approximately scale with the cosmic ray ionization rate as (McCall et al. 2003):

$$n(H_3^+) = \left(\frac{\xi}{k_e}\right)\left[\frac{n(H_2)}{n_e}\right] \quad (cm^{-3}) \qquad (3)$$

where $k_e$ is the dissociative recombination rate of $H_3^+$. Equation 3 shows that the predicted $H_3^+$ abundance will scale linearly with $\xi$. Current estimates of $\xi$ for the galactic background put it at ~$5\times10^{-17}$ $s^{-1}$ (Williams et al. 1998). However, $\xi$ can vary from sightline to sightline (see Shaw et al. 2008 and references therein). A recent estimate towards ζ Persei (Indriolo et al. 2007) finds a value of $\xi$ which is 10 times higher than the galactic background, and also mentions that $H_3^+$ should be a reliable tracer of $\xi$ in dense molecular environments. Pellegrini et al. (2006) also found enhanced $\xi$ in M17, a denser star-forming region. Clearly, the cosmic ray ionization rate needs to be well known, or the robustness to changes in $\xi$ must be explored, when constructing a PDR model.

## 3 A Series of Calculations

We present a series of calculations to show how the results of a single, time steady PDR model ("single" meaning fixed values for $n_H$ and $G_0$) change with different treatments of various physical processes. In reality, there are many more parameters that could be varied. Therefore, for conciseness, we limit ourselves to the physical processes mentioned in Section 2.

Our calculations use version C06.02 of the spectral synthesis code Cloudy[1] (Ferland et al. 1998). Cloudy is a 1D plasma simulation code which can treat the physical processes of a wide variety of astrophysical environments, such as $H^+$ regions, PDRs, and molecular clouds. Cloudy treats all energy regimes, resolving the incident radiation field into a large number of cells. All atomic photo processes, along with $H_2$ photo processes, are calculated by integrating the product of the incident radiation field intensity over the cross section for the photorate. Cloudy solves for the ionization structure for all stages of ionization for the lightest 30 elements.

PDR physics in Cloudy has recently been discussed in Abel et al. (2005), Shaw et al. (2005), and van Hoof et al. (2004). Cloudy was part of a PDR comparison study performed in Leiden, The Netherlands. This workshop compared the chemical and thermal structure along with the predicted spectrum of many PDR codes. All groups at the workshop agreed to a set of input model parameters and microphysical assumptions. Given these constraints, Cloudy agreed very well with other PDR codes (Röllig et al. 2007).

---

[1] More information about Cloudy can be found at www.nublado.org

Our goal is to show how different treatments of important physical processes in a PDR can change model predictions. We accomplish this by setting up a "standard model". We then study variations in the model predictions by tweaking the treatment of a single physical process. In the calculations presented below, we assume no hydrogen ionizing radiation. This assumption is common for PDR models, and holds true for regions such as reflection nebulae. PDR calculation results can change with the presence of an adjacent $H^+$ region (Abel et al. 2005; Kaufman et al. 2006).

## 3.1 The Standard Model

### 3.1.1 *Abundances*

Our assumed abundances are consistent with estimates for the average abundances in the interstellar medium (ISM). We consider the lightest 30 elements in our models. We take our abundances from the work of Cowie & Songalia (1986) for the warm and cold phases of the interstellar medium and Table 5 for the warm and cool phases towards ξ Oph (Savage & Sembach 1996). The oxygen abundance is from Meyer et al. (1998). For the most important species, the abundances by number are He/H = 0.098, C/H = $2.5 \times 10^{-4}$; O/H = $3.2 \times 10^{-4}$, N/H = $8 \times 10^{-5}$, Ne/H = $1.2 \times 10^{-4}$, Ar/H = $2.8 \times 10^{-6}$, and S/H = $3.2 \times 10^{-5}$. We use $n_H = 10^4$ cm$^{-3}$ in all calculations.

### 3.1.2 *Photo- and Cosmic-ray Rates*

To simplify our calculations, we use the Draine (1978) UV radiation field. Our calculations assume the integrated intensity between 11.2 – 13.6 eV relative to the average value in the ISM ($\chi$) of $10^4$. The relationship between the Draine field and $G_0$ is $1G_0 = 1.71\chi$ (Draine & Bertoldi 1996). We do not include any IR component due to hot dust from an H II region, but do calculate the UV absorption leading to IR re-emission due to dust. This incident continuum is similar to the radiation field assumed in the Leiden PDR comparison study (Röllig et al. 2007).

Our standard model calculation of photorates differs in important ways from equation 2. All atomic photoionization rates are calculated by integrating the cross section for photoionization over the local radiation field, including locally emitted line radiation such as Lyα. Most molecular photoionization and photodissociation rates use equation 2, with the parameters for $\alpha$ and $\beta$ taken from the UMIST database (Le Teuff, Millar, & Markwick, 2000). Perhaps the biggest difference is our treatment of $H_2$ self-shielding, which we calculate by considering various state-specific formation and destruction processes, solving for all 301 levels within the $H_2$ ground electronic state, and all levels within the lowest six electronic excited states. More details are given in Shaw et al. (2005). For CO, we use the self-shielding formula given in Hollenbach, Tielens, & Takahashi (1991).

The other ionization source we consider is cosmic rays. We include primary and secondary cosmic ray ionizations as described in Appendix C of Abel et al. (2005). For the standard model, we use a $\xi$ (for $H_2$) of $5\times10^{-17}$ s$^{-1}$.

### 3.1.3 *Dust Physics*

The grain physics is described in van Hoof et al. (2004). We self-consistently determine the grain temperature and charge as a function of grain size and material, for the local physical conditions and radiation field. This determines the grain photoelectric heating of the gas, an important gas heating process, as well as collisional energy exchange between the gas and dust. We also treat stochastic heating of grains as outlined in Guhathakurta & Draine (1989), which can increase the shorter wavelength grain emission. We include grain charge transfer as a general ionization – recombination process for atoms only, as described in Appendix B of Abel et al. (2005). The rates at which $H_2$ forms on grain surfaces is derived using the temperature- and material-dependent rates given in Cazaux & Tielens (2002).

The assumed grain size distribution is representative of the general ISM. We use the size distribution given in the first entry in Table 1 of Weingartner & Draine (2001b) consisting of graphites and silicates, with each species resolved into 10 size bins. Such a grain size distribution approximates the $R_V$ = 3.1 extinction curve characteristic of the ISM. The grain abundance is scaled such that $A_V/N(H_{tot})$ = $6.3\times10^{-22}$ mag cm$^2$, as used in the chemical models of van Dishoeck & Black (1986, 1988). The total grain volumes of graphite and silicates used were $2.46\times10^{-27}$ and $2.84\times10^{-27}$ cm$^3$ H$^{-1}$, respectively. Absorption and scattering efficiencies (*Q* factors) for both graphite and silicate grains are taken from Martin & Rouleau (1991). The optical properties for silicates and graphites do differ somewhat from those used in Weingartner & Draine (2001b). For silicates the difference is small, but for graphites some differences arise for the smallest grain sizes, since we do not use Li & Draine (2001) for the smallest carbonaceous grain populations.

We include size-resolved PAHs in our calculations. The size distribution for PAHs is given by a power law of the form $a^{-3.5}$, where *a* is the PAH radius and the number of carbon atoms range from 30 to 500. Just like for the silicates and graphites, we distribute the PAHs into 10 bins. The abundance of carbon atoms in PAHs relative to hydrogen, $n_C(PAH)/n_H$, is chosen to be $3\times10^{-6}$. This number is based on studies of Orion, where the elemental carbon abundance is ~$3\times10^{-4}$ (Baldwin et al. 1996) and the fraction of C in the form of PAHs is ~1% (Allamandola et al. 1989). PAHs are thought to be destroyed by hydrogen ionizing radiation and coagulate in molecular environments (see, for instance, Omont 1986). We heuristically model this effect by scaling the PAH abundance by the ratio of $H^0/H_{tot}$. The explicit abundance is $n_C(PAH)/n_H = 3\times10^{-6} \times [n(H^0)/n(H_{tot})]$.

Condensation of CO, $H_2O$, and OH onto grain surfaces is also considered following the formalism of Hasegawa, Herbst, & Leung (1992), Hasegawa & Herbst (1993), and Bergin et al. (1995). For each molecular species, we balance condensation on grain surfaces with desorption due to thermal and cosmic ray evaporation. We do not consider grain surface reactions between molecules. The effect of molecular condensation can be significant in regions where the gas and dust temperature fall below 20-25 K (Bergin et al. 1995). Removing a molecule from the gas phase will alter many of the reaction rates and will cause the molecular abundances to change from the case of pure gas-phase chemistry.

### 3.1.4 *Geometry, Turbulence, and Stopping Condition*

We assume a constant density, plane-parallel slab geometry illuminated from one side. This geometry was chosen to match that used in Tielens & Hollenbach (1985). Other geometries, such as double-sided illumination (see van Dishoeck & Black 1986) or spherical geometry (Doty & Neufeld 1997) is also commonly employed, but overall we chose this geometry for simplicity.

Our calculations include a small turbulent linewidth $\Delta v_{turb} = 1$ km s$^{-1}$. The effects of $\Delta v_{turb}$ are to reduce the predicted line optical depth, since $\tau$ and linewidth are inversely proportional, and (if the calculation were isobaric) to add an extra pressure term to the equation of state. We do not explore variations in this parameter.

The final parameter in our standard model is the stopping criterion. We stop our calculation when $A_V = 100$ mag (for a point source). We chose this value in order to reach conditions cold enough for an appreciable fraction of molecules to freeze-out on grain surfaces.

## 3.2 Physical Processes Allowed to Vary

Given the parameters of the standard model, we performed a series of calculations in which we varied a single physical process. In this section, we define each tweaked model with a roman numeral.

### 3.2.1 *Dust Physics*[1]

I. Instead of ISM dust with $R_V = 3.1$, we use the $R_V \sim 5.5$ distribution defined by the 8th entry (case A, $b_C = 0$) in Table 1 of Weingartner & Draine (2001b). This value of $R_V$ is characteristic of grains in Orion.

II. Instead of ISM dust with $R_V = 3.1$, we use the $R_V = 4$ distribution defined by the 13th entry (case B, $b_C = 0$) in Table 1 of Weingartner & Draine

---

[1] All models are constrained to have the same $A_V/N(H_{tot})$ ratio of $6.3\times10^{-22}$ mag cm$^2$

(2001b). This value of $R_V$ is characteristic of starburst galaxies (Calzetti 2001). Overall, the standard model, along with models I & II, represent lower and upper limits to $R_V$ and therefore the effects of plausible ranges of grain sizes in star-forming regions.

III. We use equations 42, 43 of Bakes & Tielens (1994) for the grain heating, instead of Weingartner & Draine (2001b).

IV. This model does not consider grain-atom/ion charge transfer. Therefore, this model is similar to the studies performed by Bakes & Tielens (1998) and Boger & Sternberg (2006) in understanding the importance of grain-atom charge transfer.

V. This model does not consider freeze out of CO, $H_2O$, and OH.

### 3.2.2 *Rates & Continuum*

VI. In this model, we increased $\xi$ by a factor of 10 over the standard model, to $5\times10^{-16}$ s$^{-1}$.

VII. A calculation where we use the radiative and dielectronic recombination rates calculated by Badnell[1] (2005), instead of the rates used in the standard model (Pequignot et al. 1991).

VIII. Instead of using the Draine (1978) radiation field, we use a Kurucz (1991) O star continuum with $T^* = 35,000$ K, normalized to the same integrated flux between 6 – 13.6eV, while also not including any hydrogen-ionizing radiation.

# 4 Results

In this section, we show how changing the treatment of various physical processes as outlined in sections 3.2.1 and 3.2.2 changes the predictions of the standard model. Our results are presented in the order given in Section 3.2.

## 4.1 Models I & II

The differences that arise because of differences in the assumed grain size distribution are not negligible. As the smaller grains are removed, the total heating rate decreases, as discussed in Bakes & Tielens (1994). Figure 1 shows $T_g$ (A) and the photoelectric heating rate (B) as a function of $A_V$. For the large Orion grains, the photoelectric heating rate is ~15% lower than the ISM grains, and about 10% lower than the $R_V = 4$ grain distribution. This leads to temperatures at the illuminated face of the PDR of 265, 220, and 240 K for the standard model, Model I, and Model II, respectively. The effects of extinction are also seen in Figure 1, as the larger Orion grains, which have a lower PDR temperature at the illuminated face, also lead to hotter PDR temperature at larger $A_V$, since the UV is extinguished less efficiently by the $R_V$ = 5.5 grain size distribution. So, while the ISM grains lead to more heating for $A_V$ < 1 mag, the Orion and $R_V$ = 4 grain size distribution leads to a hotter PDR for $A_V$ = 1 – 10 mag. Once grain heating becomes insignificant, other processes such as cosmic ray heating determine $T_g$ and all grain size distributions converge to a common temperature of ~13 K.

---

[1] Atomic data can be found at http://amdpp.phys.strath.ac.uk/tamoc/DATA/DR

Differences in the grain-size distribution also affect the predicted column densities shown in Figures 2 & 3. For $H_2$ (Figure 2A), the predicted column densities can vary by 1 - 4 orders of magnitude, for $A_V$ between 0.7 – 1.5 mag. This is again due to the dependence on grain size in extinguishing UV radiation. While the range in $N(H_2)$ which is sensitive to $R_V$ are typically not observed for the combination of $n_H$ and $G_0$ considered here, $R_V$ is almost certainly important in lower density, lower $A_V$ translucent clouds, as shown in Abel et al. (2004). For $R_V$ = 4 and 5.5 grains, the UV radiation is extinguished at less efficiently, thereby increasing the $H_2$ photodissociation rate over a larger physical extent. The UV extinction properties of the grains also affects the $C^+$/C/CO transition (Figures 3A – C), as deeper penetration of UV radiation allows for a large $C^0$ photoionization rate (Figure 3D) over a longer path length. This leads to more $N(C^+)$ in the Orion and Starburst grain models. This also moves the formation of C and CO at larger $A_V$. The general effects of $R_V$ on the $C^+$ region shown here are also discussed in Tielens & Hollenbach (1985). Since $H_2$ and the carbon chemistry are vital to the chemistry in a PDR, other molecules are also affected, with large differences arising in $N(OH)$ and $N(H_3^+)$ for $A_V$ ~1 – 20 mag (Figures 2B - C).

## 4.2 Model III

Figure 4 shows the differences in the photoelectric heating rate and temperature that arise when the heating rate of Bakes & Tielens (1994 –BT94) is used. Figure 4 compares the heating rate and temperature derived from the microphysics, using an ISM grain distribution, to BT94. The BT94 heating rate is a factor of 3 higher than the standard model, which leads to a much hotter PDR at the surface than for the standard model (~900 K vs. 265 K).

Figure 15 of WD01 shows the heating rate at the PDR surface from BT94 and for various other grain size distributions, versus the parameter $\dfrac{G_0\sqrt{T_g}}{n_e}$. For the standard model, $\dfrac{G_0\sqrt{T_g}}{n_e} = 2\times10^5$ K$^{1/2}$ cm$^3$ for BT94 and $10^5$ K$^{1/2}$ cm$^3$ for $R_V$ = 3.1. Figure 15 shows that, for these values of $\dfrac{G_0\sqrt{T_g}}{n_e}$, the BT94 heating rate should be a factor of 3-4 larger than the $R_V$ = 3.1 grain size distribution, (the grain model used here corresponds to $b_C = 0$). This agrees well with our Figure 4b. Overall, our results point to the sensitivity of the heating rate to the grain size distribution (as shown in Weingartner & Draine 2001a) and how grain size distributions that reproduce the same value of $R_V$ can still lead to significantly different heating rates.

The higher temperature given by the BT94 rate also affects molecular abundances. As an illustration, we show $N$(OH) vs. $A_V$ for the standard model and for BT94 heating (Figure 4C). The hotter temperature predicted by BT94 initiates the O + $H_2$ → OH + H reaction, which has a temperature barrier ($c$ in equation 1) of 3160 K. The difference in $T_g$ between BT94 and WD01 increases the rate of this reaction by 3 orders of magnitude, which leads to a 1 dex increase in the predicted column density for an $A_V$ of 1-4 mag.

## 4.3 Model IV

Figure 5 shows the effects of disabling grain charge transfer on the chemistry. The primary effect of grain-atom/ion charge transfer is to lower the electron abundance by acting as a catalyst to bring electrons and heavy element ions together (Bakes & Tielens 1998). In our model, $n_e$ is lowered by > 1 dex when grain charge transfer is included. The >1 dex increase in $n_e$ has a major effect on the predicted molecular column densities for species which are created/destroyed through electron recombination reactions. Figure 5 shows $N(H_3^+)$, $N$(OH), and $N(O_2)$ versus $A_V$. Increasing $n_e$ causes a decrease in the $H_3^+$ abundance, since $H_3^+$ is destroyed through electron recombinations (see also equation 3). Increasing $n_e$ also produces more OH through the e + $H_2O^+$ and e + $H_3O^+$ reaction channels. The increased OH abundance increases the $O_2$ abundance through the O + OH → $O_2$ + H reaction (Bergin et al. 2000). Overall, not including grain charge transfer in the chemistry leads to 1 dex or higher changes in the predicted molecular column densities.

The difference in $N(H_3^+)$ is particularly important for models which deduce $\xi$. Equation 3 shows that the $H_3^+$ scales linearly with $\xi$ and inversely with $n_e$. A model for a region with an observed $H_3^+$ density that does not include grain charge transfer will deduce a larger $\xi$ than a model with grain charge transfer included (Shaw et al. 2008). The predicted abundance of interstellar molecules is clearly sensitive to the coupling between $n_e$, $\xi$, and grain charge. This coupling is also important in eliminating the bi-stability of chemical models, as discussed in Boger & Sternberg (2006).

## 4.4 Model V

Figure 6 shows the effects of disabling the freeze-out of molecules on grain surfaces. Shown are the abundances of OH (A) and $O_2$ (B). With freeze-out included, once the gas/dust temperature gets cold enough (~20 K), a significant fraction of $H_2O$ becomes solid. Removing water from the gas phase alters the chemistry by reducing the gas-phase oxygen abundance and closing off formation pathways to various molecules initiated by gas phase $H_2O$. For this particular model, $N(OH)$ is a factor of 2 lower and $N(O_2)$ about 5 times smaller with freeze-out included.

Recent modeling by Kaufman et al. (2005) shows the need to include freeze-out in PDR calculations. The results of SWAS and ODIN show that $O_2$ is much less abundant than predicted by pure gas-phase chemistry. Kaufman et al. (2005) show that a possible explanation of this is the effects of $H_2O$ freeze-out on the molecular abundances. Our model confirms this.

## 4.5 Model VII

Figure 7a and b shows the differences that arise between the predicted $N(C^0)$ when the radiative and dielectronic recombination rates of Badnell (2005) are used instead of the rates of Pequinot et al. (1991--radiative) and Nussbaumer & Storey (1983-dielectronic). For the Nussbaumer & Storey (1983) rates, since they are only defined for $T > 1000$ K, we used the low-$T$ cutoff times a Boltzmann factor appropriate for the energy of the auto ionizing level. For $A_V < 3$ mag, $N(C^0)$ is about 30% higher when the Badnell rates are used. This is due to the Badnell rates having a larger $C^+$ recombination rate at a given $T_g$, which leads to a larger $C^0$ formation rate. For $A_V > 3$ mag, $C^+$ recombination is no longer the dominant formation process for $C^0$, instead chemical reactions whose by-product is $C^0$ become important. This is reflected in Figure 7a, as both the Badnell rate and the standard model predict nearly identical $N(C^0)$, even though the $C^+$ recombination rate is significantly higher for the Badnell rates.

The differences in $N(C^0)$ due to differences in the recombination rate are potentially very important. The primary differences between the Badnell rates and the previous recombination rates are that the dielectronic recombination rates are much higher than in Nussbaumer & Storey (1983). Use of these rates show that PDR models of regions where CO has not fully formed can under-predict $N(C^0)$ by about 30%, which could potentially affect the FIR [C I] emission-lines which should be commonly observed by Herschel and SOFIA.

## 4.6 Model VIII

Figure 8 shows the predicted carbon photoionization rate and $T_g$ for the standard model and a 35,000 Kurucz (1991) continuum with an identical FUV flux between 6 – 13.6eV, but no H-ionization radiation. Figure 9 shows the shape of the incident continuum and the continuum when $A_V = 1$ mag, which corresponds to the depth into the cloud where the difference between the predicted Tg for both models is the largest. The Kurucz continuum has a larger flux over C-ionizing energies, which leads to a factor of two increase in the $C^0$ ionization rate for the Kurucz (1991) continuum, for $A_V < 4.5$ mag. The predicted $T_g$ is higher for the Kurucz continuum for $A_V$ between 0.3 – 2 mag. This is due to an increase in the grain photoelectric heating rate in this regime, which arises due to more photons with energies >10.2 eV (Figure 9).

Differences in PDR heating due to differences in the shape of the incident continuum have been discussed elsewhere. Spaans et al. (1994) found the ratio of $I_{[CII]} / \sum_{\text{all lines}} I_{CO}$ to be a good diagnostic of the radiation fields color temperature. Our choice for the continuum shape is not as extreme as in Spaans et al (1994), who studied color temperatures ranging from a 6,000 to 30,000 K blackbody. Our comparison is for roughly the same color temperature (30,000 vs. 35,000 K), but with different continuum shapes. As shown in Section 5.1.4, differences in the continuum shape while keeping the 6 – 13.6eV flux fixed can have a significant effect on the $H_2$ spectrum.

# 5 Effects on the PDR emission-line spectrum

Perhaps the most interesting results of our calculations are the differences in the predicted emission-line spectra that arise by changing the treatment of a single physical process. Figures 10-14 show the predicted [C II], [O I], and [C I] infrared fine-structure emission for each model. Figure 15, 16, and 17 show the $H_2$ $0-0$ $S(1)$ 17.03μm / $0-0$ $S(0)$ 28.2μm, $H_2$ $0-0$ $S(2)$ 12.28μm / $0-0$ $S(0)$ 28.2μm, and $H_2$ $0-0$ $S(3)$ 9.66μm / $0-0$ $S(0)$ 28.2μm emission-line ratios commonly observed with the Spitzer Space Telescope. Each of these spectral features shows some variation with changes in the treatment of physical processes, as described below.

### 5.1.1 *[C II] 158μm*

Figure 10 shows the predicted [C II] 158 μm intensity (henceforth $I_{[C\,II]}$). [C II] emission is efficiently produced by regions with $T_g$ > 92 K, with the upper level of the $C^+$ transition located 92 K above the ground state (see Kaufman et al. 1999). We find that the biggest changes in $I_{[C\,II]}$ are due to changes in the grain physics. Figure 1A shows that, compared to the standard model, both the Orion and Starburst grain models produce a slightly larger region with temperatures above 92 K. This leads to a ~15% increase in $I_{[C\,II]}$ for the Orion/Starburst grain models. The larger $N(C^+)$ (see Figure 3B) for Model I and II also contributes to increased emission. The single largest increase in [C II] emission occurs when the BT94 heating rate is used, which efficiently heats the gas to temperatures of ~$10^3$ K, and also produces a larger $T_g$ > 92 K region than the standard model. This leads to a 50% increase in $I_{[C\,II]}$ over the standard model. $N(C^+)$ for the BT94 model is within 5% of the standard model, so the differences between $I_{[C\,II]}$ for the BT94 and standard model are entirely due to differences in temperature. There is also a ~5% decrease in $I_{[C\,II]}$ when the Kurucz continuum is used due to the slight differences in heating and photoionization rates for $C^0$ which arise from the different UV continuum shape.

### 5.1.2 *[O I] 63, 146 μm*

Figures 11 & 12 show the variation in $I_{[O\,I]\,63,\,146\,\mu m}$. The upper levels for the 63, 146 μm lines are located 228 and 327 K above the lowest level. Again, we find the most significant variation is due to variations in grain physics. Again, since the Orion and Starburst grain models produce a hotter PDR over a larger region, [O I] emission is somewhat higher for Model I and II over the standard model. This explains why these two models produce significantly more [O I] 63 μm emission. The [O I] 63 μm emission is a factor of 30% higher for $R_V = 4$ grains and ~50% higher for Orion grains compared to the standard model, with similar differences for $I_{[O\,I]\,146\,\mu m}$. The BT94, again owing to the hotter temperature, produces the most [O I] emission. Since the temperature needed to produce [O I] is greater, the dependence on temperature is even more sensitive, with $I_{[O\,I]\,63\,\mu m}$ being a factor of 4 higher and $I_{[O\,I]\,146\,\mu m}$ almost a factor of 5 higher than the standard model. The cooler Orion grain model is less efficient in populating the $^3P_0$ level, therefore $I_{[O\,I]\,146\,\mu m}$ was ~10% lower compared to the standard model.

### 5.1.3 *[C I] 370, 610 μm*

The FIR [C I] emission lines (Figures 13 and 14) show the most sensitivity to changes in the model parameters. The upper levels of the [C I] 370, 610 μm lines are at ~62 and ~24 K, respectively. Again, grain physics plays an important role in changing the predicted intensities. We find that as the average grain size is increased, $I_{[C\,I]\,370,\,610\,\mu m}$ increases. This is due to the larger grains producing a larger region where $T_g$ is high enough to efficiently excite the [C I] levels. Additionally, the reduced efficiency of the larger grains to extinguish the UV leads to a larger $N(C^0)$ for Model I & II. This can be seen by comparing Figures 1A and 3A, which shows the value of $N(C^0)$ at the depth where $T_g$ falls below 24 and 62 K, and the relative emission from the ISM ($R_V = 3.1$), Starburst ($R_V = 4$) and Orion ($R_V = 5.5$) grain models are strongly correlated. The combined effects lead to Orion grains producing ~50% more $I_{[C\,I]\,370,\,610\,\mu m}$ emission than the ISM model, while the Starburst grains (Model II) yield about 20% more emission. The BT94 heating rate, which has the same $N(C^0)$ dependence with depth as the standard model (Figure 3A), has about the same increase in emission as the Orion grains.

For $I_{[C\,I]\,370,\,610\,\mu m}$ emission, we find that either increasing the cosmic ray ionization rate or changing the continuum source increased the emission by 15-20% relative to the standard model. There are several reasons for the increased emission. Cosmic rays are often the dominant heating agent deep in a PDR. Increasing $\xi$ therefore increases the temperature at high $A_V$. In addition, increasing $\xi$ increases the abundance of $C^0$ deep in the PDR, through destruction of CO, either directly through cosmic ray destruction or indirectly through the formation of $H_3^+$ and $He^+$, which then reacts with CO. The Kurucz continuum produces a slightly smaller $T_g$ > 24 K region than the standard model, which in turn leads to a ~15% decrease in $I_{[C\ I]\ 370,\ 610\ \mu m}$ emission.

### 5.1.4 $H_2$ Pure Rotational Emission-Line Ratios

Figures 15 – 17 show the effects of grain physics, cosmic-rays, and the SED shape on the $H_2$ emission line ratios commonly observed by Spitzer. Unlike the atomic PDR lines, which are theoretically modeled by all PDR calculations, modeling the $H_2$ spectrum requires a detailed calculation of the $H_2$ level populations, which is done by only a few PDR research groups (see Röllig et al 2007, Kaufman et al. 2006 for a summary of those groups). Since our calculations employed the full $H_2$ model atom implemented in Cloudy (Shaw et al. 2005), we have extracted this spectrum in order to see how changes in treatment of physical processes change the $H_2$ spectrum.

Our calculations show the sensitivity of the $H_2$ spectrum to the thermal structure. The $H_2$ ratios decrease with increasing grain sizes, with 30 – 60% smaller line ratios for the Orion and Starburst grain models. The Bakes & Tielens (1994) photoelectric heating rate can lead to increases in the emission –line ratios of 2 – 10 compared to the standard model. The $H_2$ ratios are much more sensitive to the shape of the SED than any of the atomic PDR emission-lines. When the Kurucz SED is used, UV pumping increases the 0-0 S(3)/ 0-0 S(0) ratio by a factor of 2 over the standard model. UV pumping of the $H_2$ lines has, of course, been known for decades, but it is instructive to show how the $H_2$ spectrum changes if you do nothing else than change the nature of the SED, and do not change the numerical value for $G_0$. Cosmic rays also have an effect on the $H_2$ spectrum, producing another non-thermal source for pumping the $H_2$ levels, as discussed recently in Shaw et al. (2008).

Our $H_2$ spectrum calculations can be compared to those presented in the recent work of Kaufman et al. (2006). Our predicted 0-0 S(1)/ 0-0 S(0) ratio varies from 6 – 16 which, when compared to Kaufman et al. (2006) Figure 8, correspond to the $n_H$ > $10^4$; $G_0$ > $10^3$ portion of the parameter space. Our predicted 0-0 S(2)/ 0-0 S(0) ratio varies between 0.3 – 4, which also corresponds to wide range in $n_H$ and $G_0$ when compared to Figure 14 of Kaufman et al. (2006).

# 6 Conclusions

The comparison given in 5.1.4 emphasizes the fundamental point of this work, a need to include variations in grain physics and UV SED shape in developing a physical model through comparison of theory and observation. As is shown with the $H_2$ spectrum, a change in the grain size distribution, grain heating model, or UV radiation field will lead to differing interpretations about the density and FUV field responsible for producing the $H_2$ spectrum. As is noted in the works of Kaufman et al. (1999, 2006), and emphasized here, only by getting many observables and performing a full investigation of all the possible physical processes that alter the spectrum can we hope to reasonably deduce physical quantities through merging theory with observation.

In this work we use the spectral synthesis code Cloudy, modified in the last five years to incorporate a wide range of PDR physical processes, to explore how the results of a PDR calculation change when the treatment of various physical processes involving the grain physics, incident continuum, and chemistry are varied. The results of our calculations show the following:

- We find an order of magnitude variation in $T_g$, without changing either $n(H)$ or $G_0$. These differences arise as a result of differences in the grain size distribution and of differences in the treatment of grain photoelectric heating. We find that different grain size distributions that yield the same $R_V$ lead to significant differences in the thermal structure, which agrees qualitatively with WD01.

- Grain charge transfer strongly affects $n_e$ deep in the PDR. When we did not consider this, $n_e$ was an order of magnitude higher. This can affect the predicted molecular column densities of OH and $H_3^+$ by an order of magnitude. When coupled with the variation in OH and $H_3^+$ that arise due to uncertainties in $\xi$, we conclude that grain charge transfer must be included in any PDR calculation where deducing $\xi$ is a goal.

- The predicted atomic and $H_2$ PDR spectrum shows significant variation, for a constant $n_H$ and $G_0$, depending on the physical process enabled/disabled. Almost all physical processes we allowed to vary affected the emission of at least one PDR line. We find the grain size distribution and heating rate to be the most important physical processes affecting the predicted spectrum, due to their effects on the temperature structure in the PDR, and due to differences in the chemical structure brought about by differences in UV extinction. However, the incident continuum shape, recombination rates, and cosmic-ray ionization rate also affected some PDR lines and the overall ionization structure.

Acknowledgements: We would like to thank the anonymous referee for many invaluable suggestions which helped to improve this manuscript. NPA would like to acknowledge the University of Kentucky Supercomputing Center, the Miami University (Ohio) Redhawk Computer Cluster, and National Science Foundation under Grant No. 0094050, 0607497 to The University of Cincinnati. NPA and GS acknowledge the Center for Computational Sciences at the University of Kentucky. PvH acknowledges support by the Belgian Science Policy Office under program MO/33/017. GJF acknowledges support by NSF (AST 0607028), NASA (NNG05GD81G), and by STScI (HST-AR-10951).

# 8 Figures

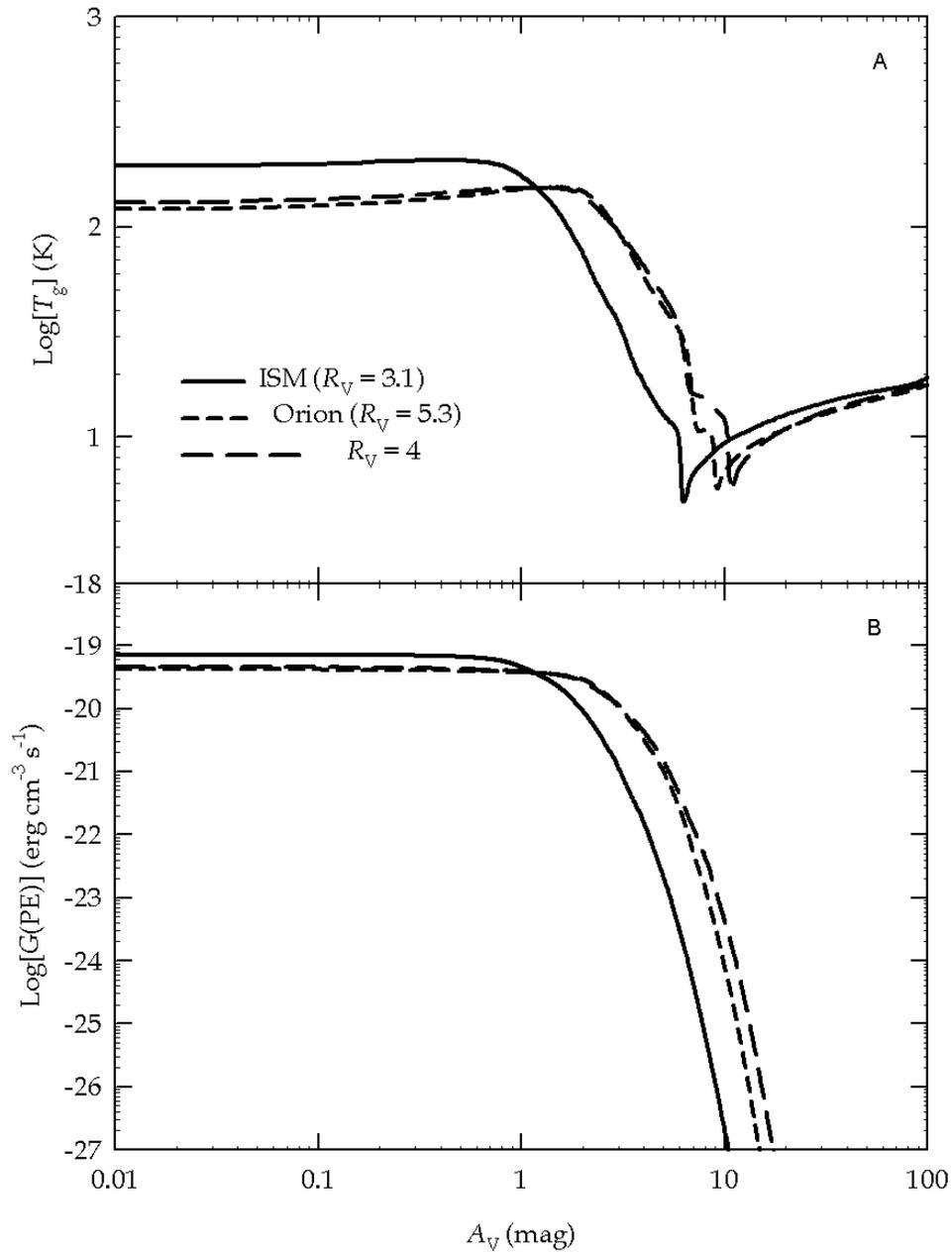

Figure 1 Gas temperature (A) and photoelectric heating rate (B) for different grain size distributions, corresponding to models I & II. Shown are the standard model (ISM, $R_V$ = 3.1), starburst ($R_V$ = 4), and larger Orion grains ($R_V$ = 5.5).

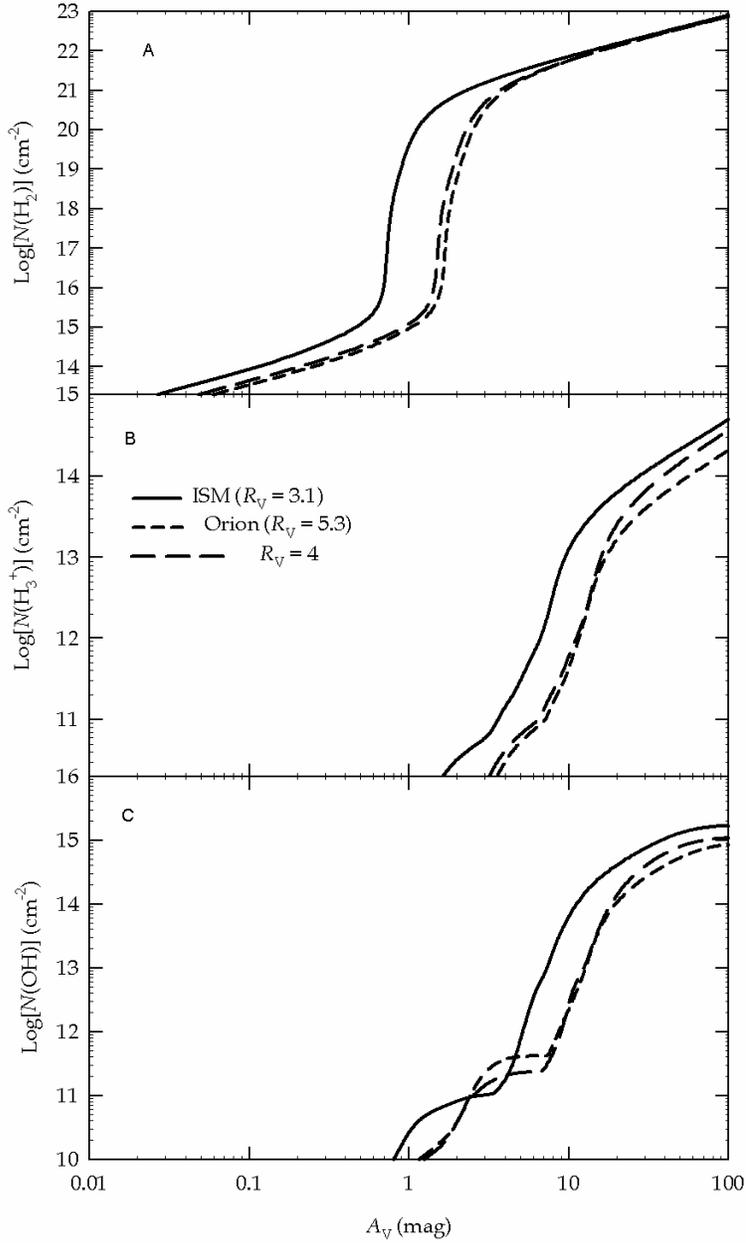

Figure 2 Predicted molecular column densities for different grain size distributions of model I & II. As the grains become larger, UV radiation penetrates deeper, destroying $H_2$ more effectively (A). Delayed $H_2$ formation, combined with lower temperatures predicted by larger grain size distributions, combine to lower the predicted column densities of other species (B, C)

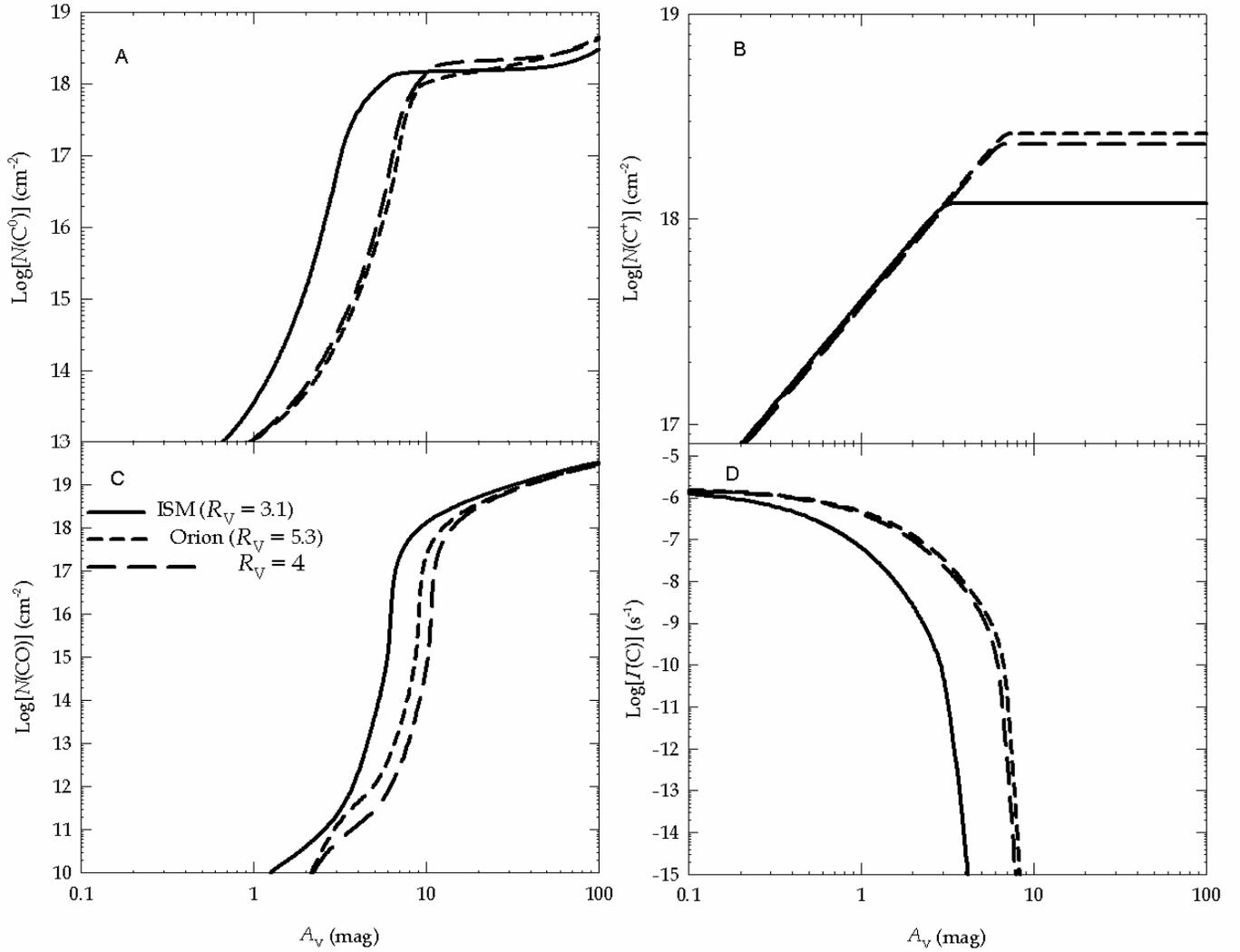

Figure 3 Predicted $C^+/C/CO$ column densities (A, B, C) along with the carbon photoionization rate for different grain size distributions given by models I &II. Larger Orion grains are less effective in extinguishing UV radiation, which translates to formation of CO at higher $A_V$ than for ISM or the starburst $R_V$ grains. Since UV radiation penetrates deeper, the $C^+$ photoionization rate is substantially higher with depth, leading to a larger $C^+$ region and $N(C^+)$ for Orion grains.

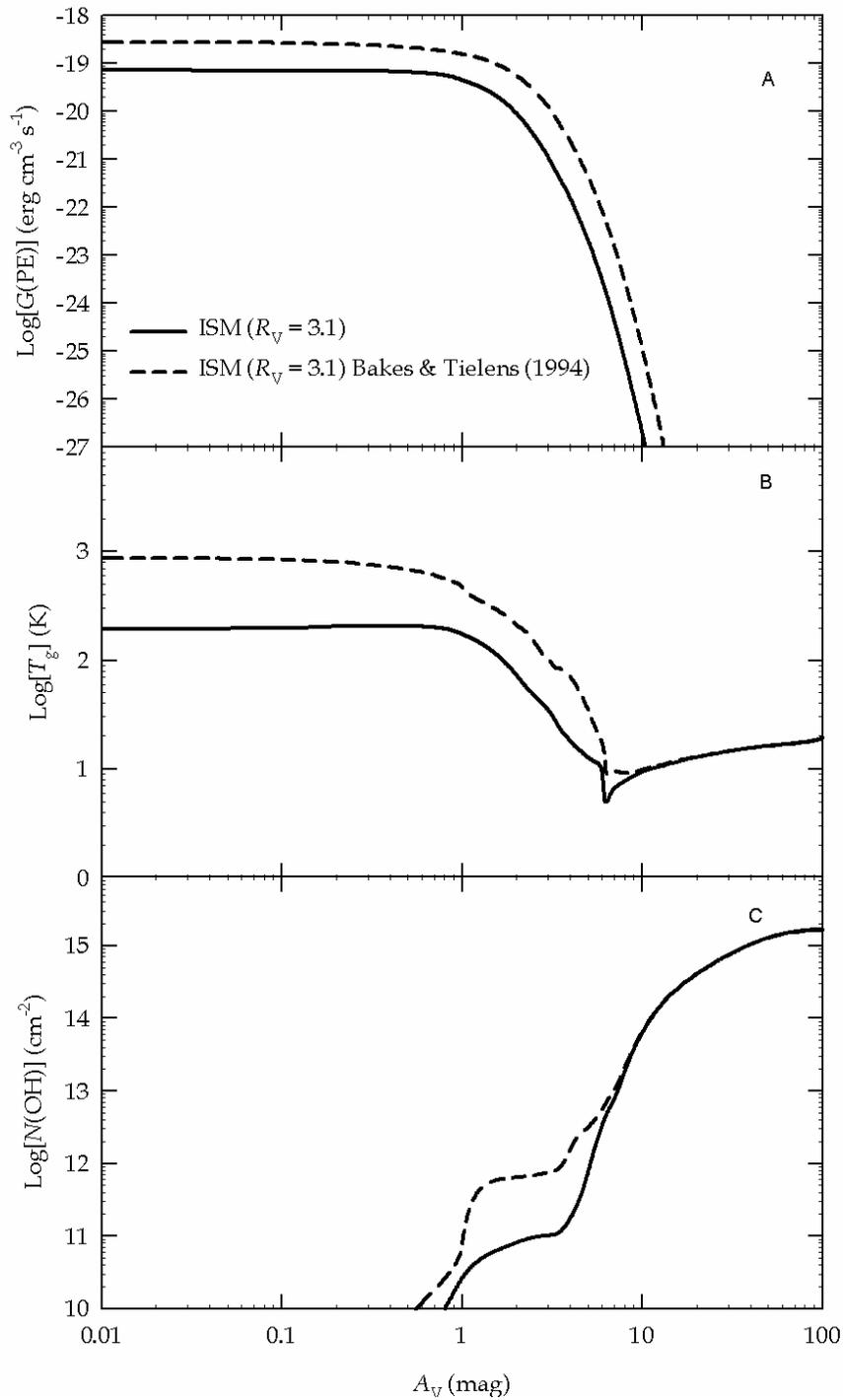

Figure 4 Photoelectric heating rate (A), Gas temperature (B), and OH column density (C) for Bakes & Tielens (1994) heating rate compared to the standard model. The BT94 heating rate is substantially larger than the standard model, leading to a warmer PDR. The higher temperatures lead to increased reaction rates for neutral-neutral reactions, such as O + $H_2$ → OH + H, which lead to more OH formation.

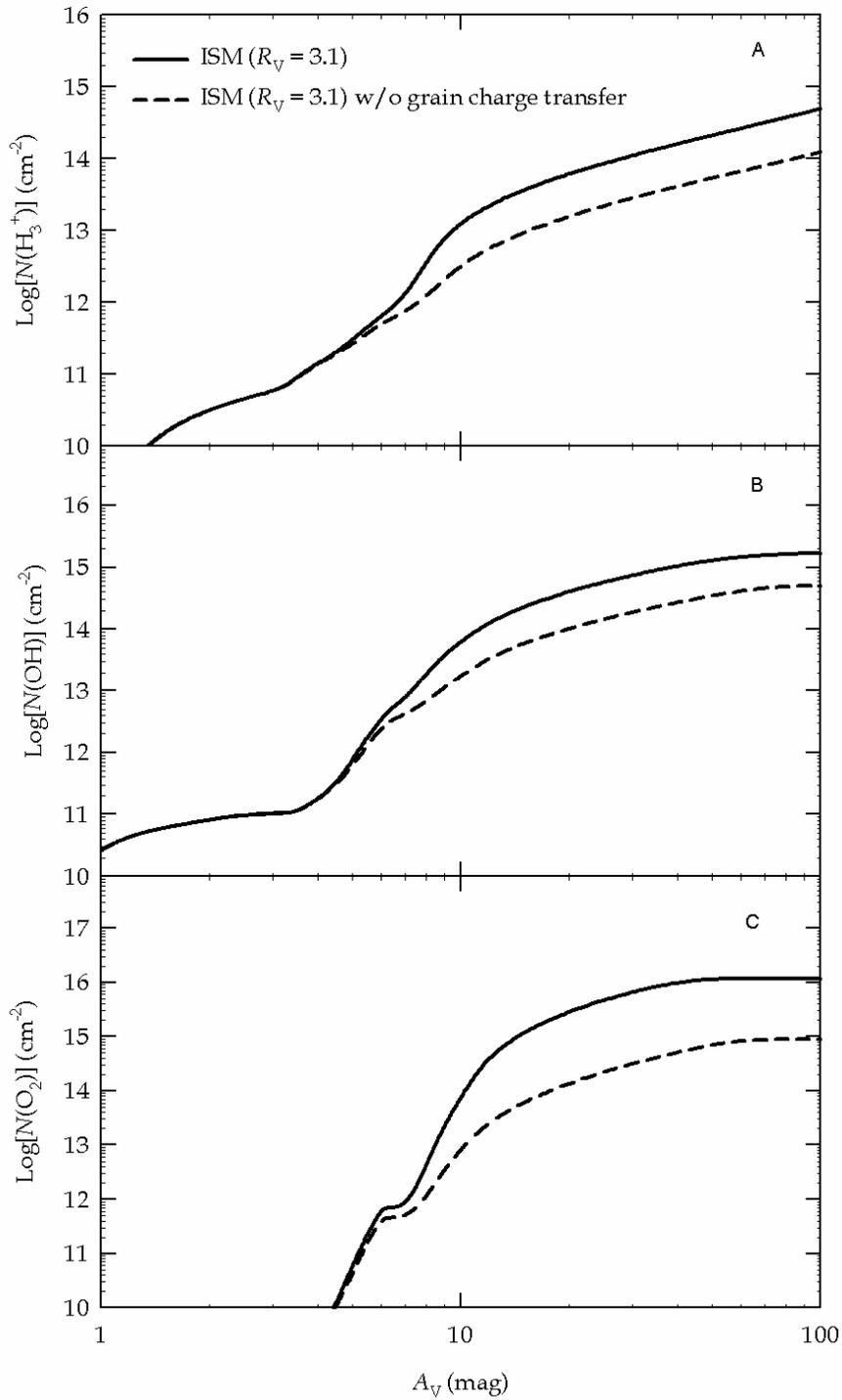

Figure 5 Molecular column densities with (corresponding to the standard model) and without grain charge transfer. Without grain charge transfer, more free electrons are produced, which through fast reactions with ionized molecules leads to significant differences in the predicted column densities.

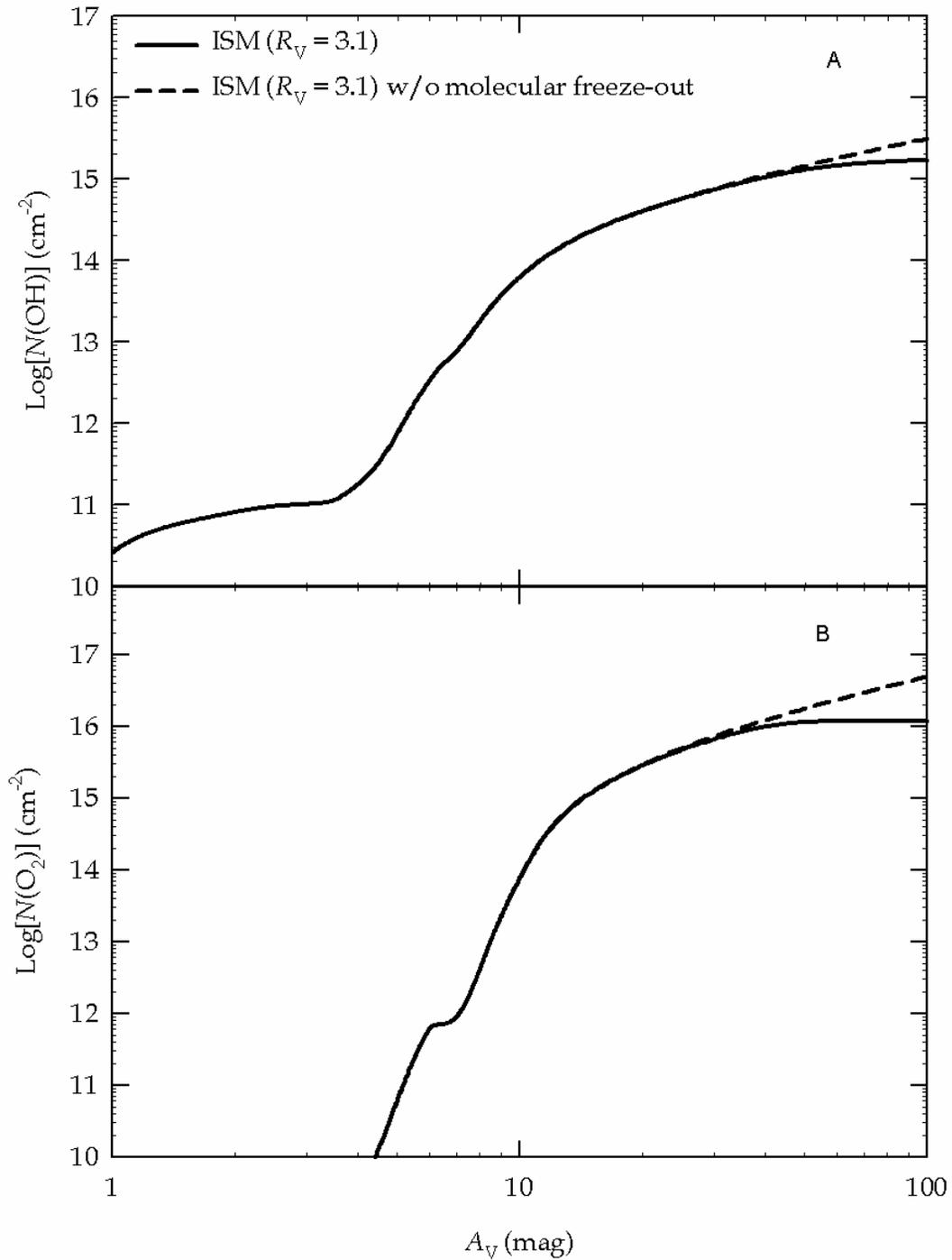

Figure 6 $N$(OH) and $N$(O$_2$), with (corresponding to the standard model) and without freeze-out of molecules onto grains. With freeze-out, as the gas and dust becomes cold, molecules can stick onto grains without thermally evaporating, which is particularly important for H$_2$O. This leads to less OH and O$_2$ by turning off various formation channels that require gas phase water. Without freeze-out, the predicted column densities are 2-4 times higher.

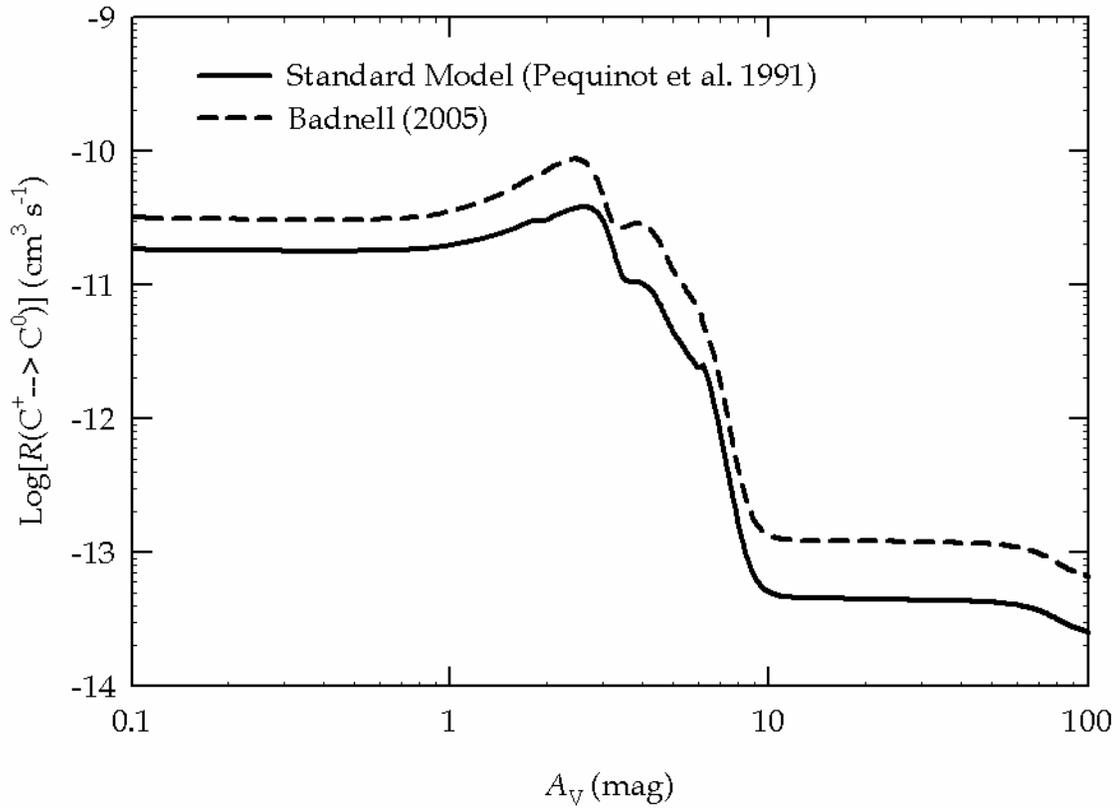

Figure 7 $C^0$ recombination rate (A) for the standard model, which uses the rates from Pequinot et al. (1991-radiative) and Nussbaumer & Storey (1983-dielectric), compared to Badnell et al. (2005), along with the predicted $N(C^0)$ (B). The Badnell et al. (2005) rates are higher at a given $T_g$, leading to more $C^0$.

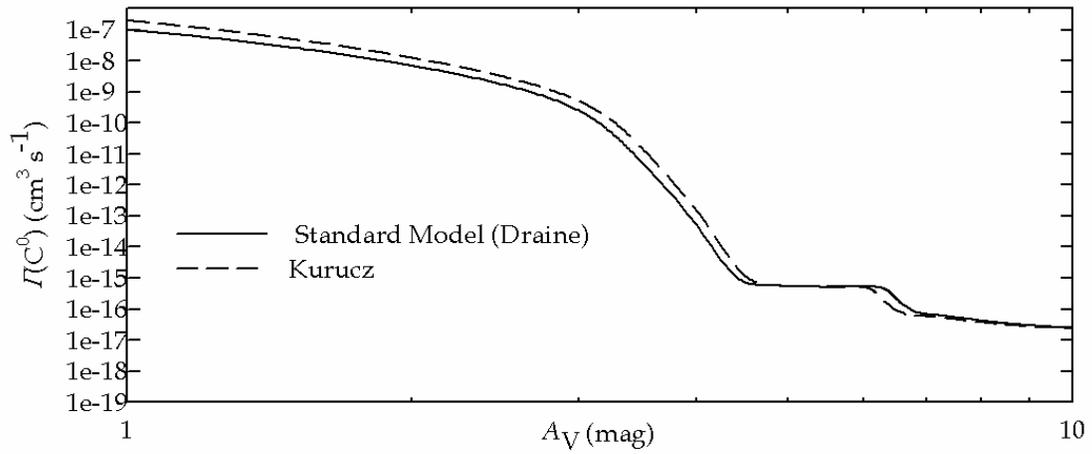

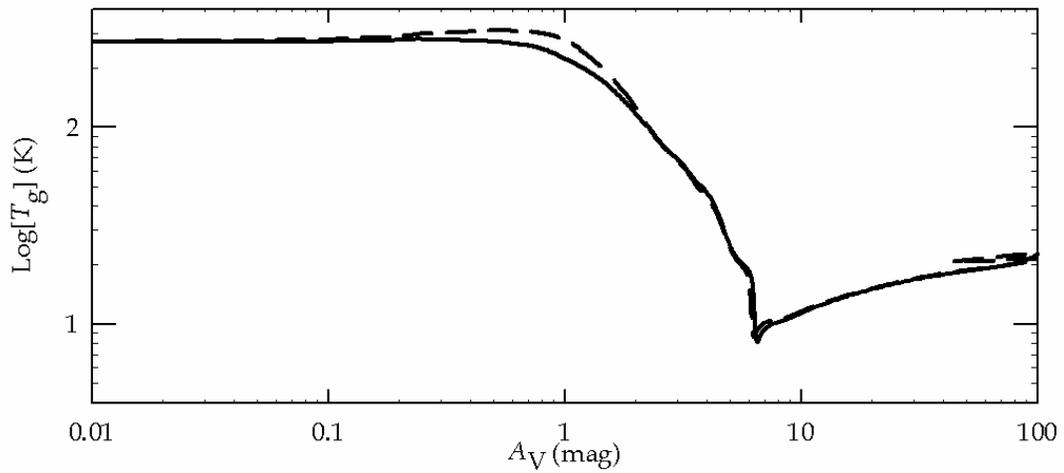

Figure 8 $\Gamma(C^0)$ (A) and $T_g$ (B) results for the standard model, which uses the Draine (1978) radiation field, and a 35,000 K Kurucz (1991) stellar atmosphere with the same $G_0$ but without the hydrogen ionizing radiation.

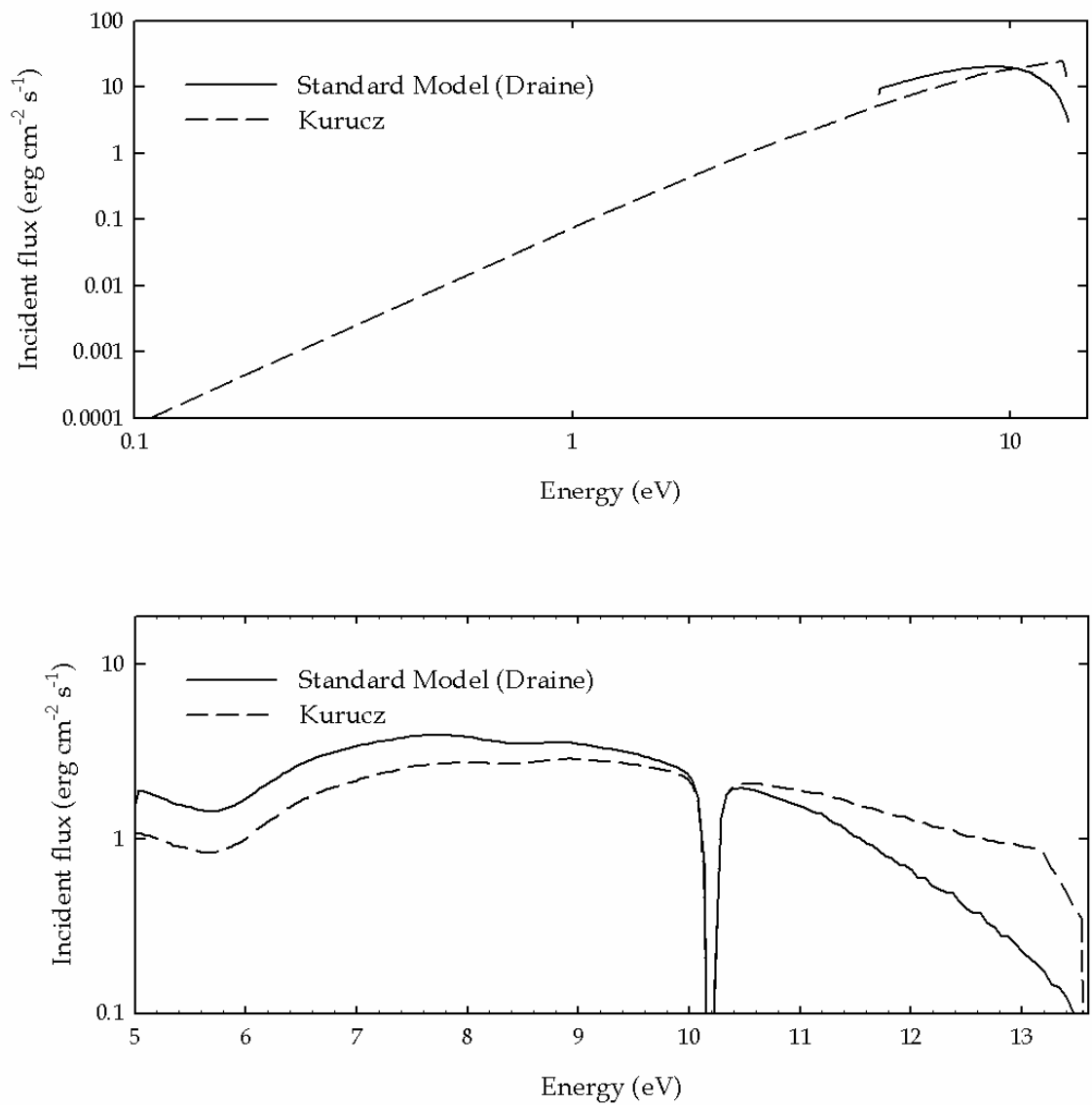

Figure 9 Incident continua for the Kurucz (1991) and Draine (1978) continuum used in our models. Both continua are normalized to have the same flux between 6 – 13.6eV. Also shown (lower figure) is the transmitted continuum between 5 – 13.6eV when $A_V$ = 1 mag. This dip at 10.2eV is due to Ly$\alpha$.

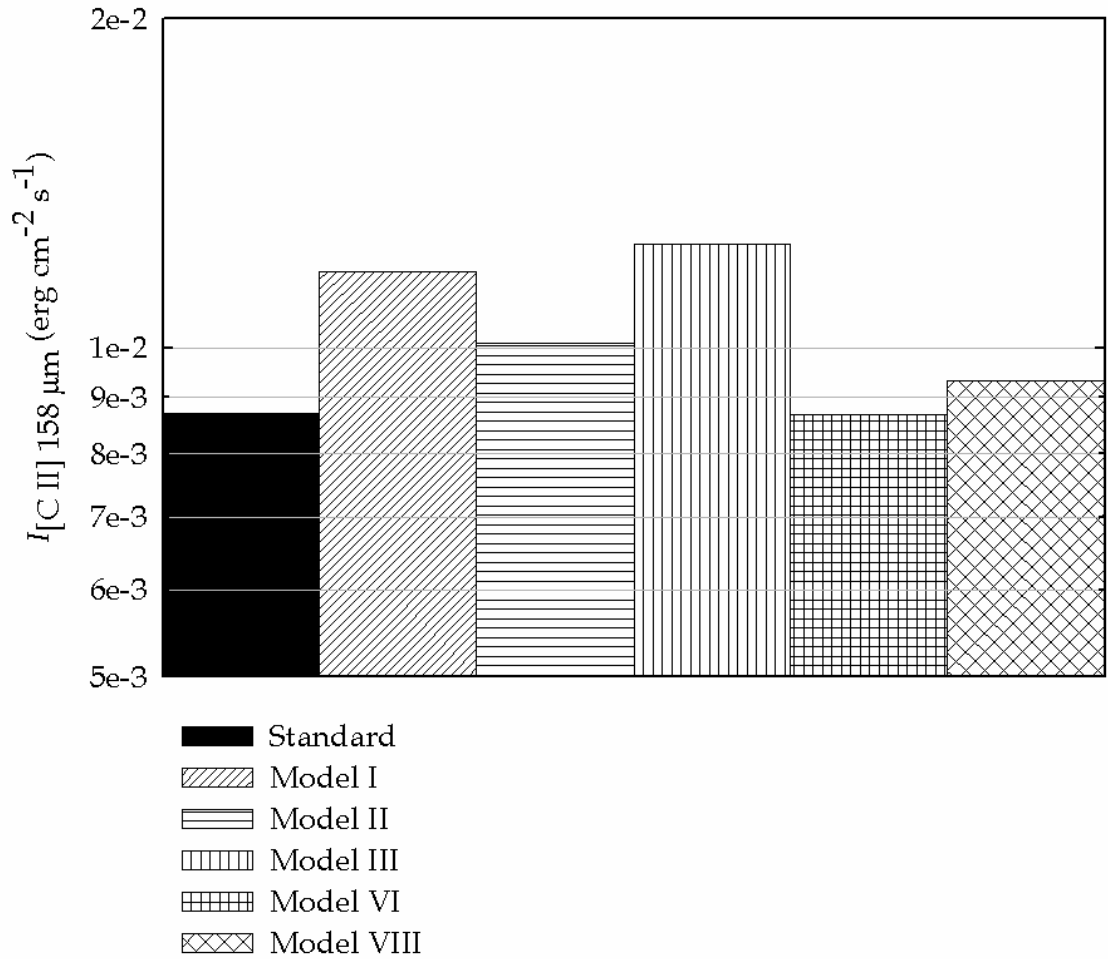

Figure 10 Predicted intensity of the [C II] 158μm line for various treatments of physical processes in a PDR.

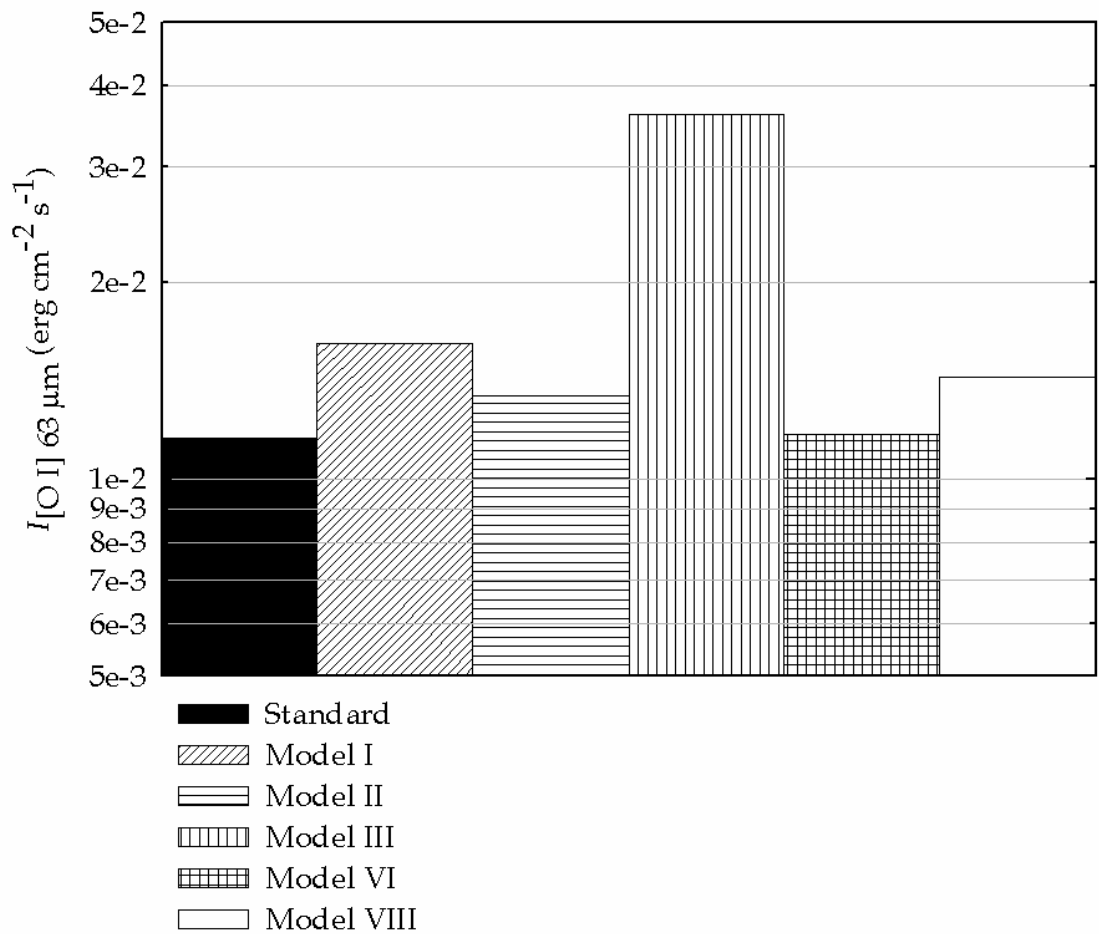

Figure 11 Predicted intensity of the [O I] 63 μm line for various treatments of physical processes in a PDR.

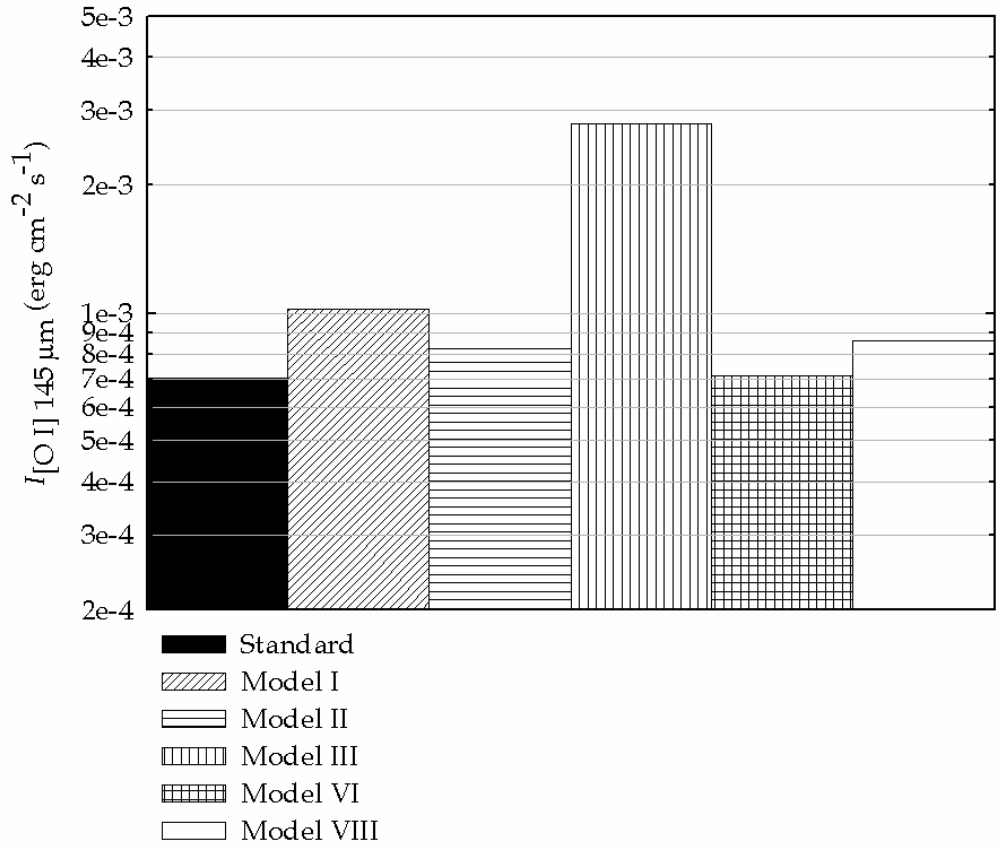

Figure 12 Predicted intensity of the [O I] 146 μm line for various treatments of physical processes in a PDR.

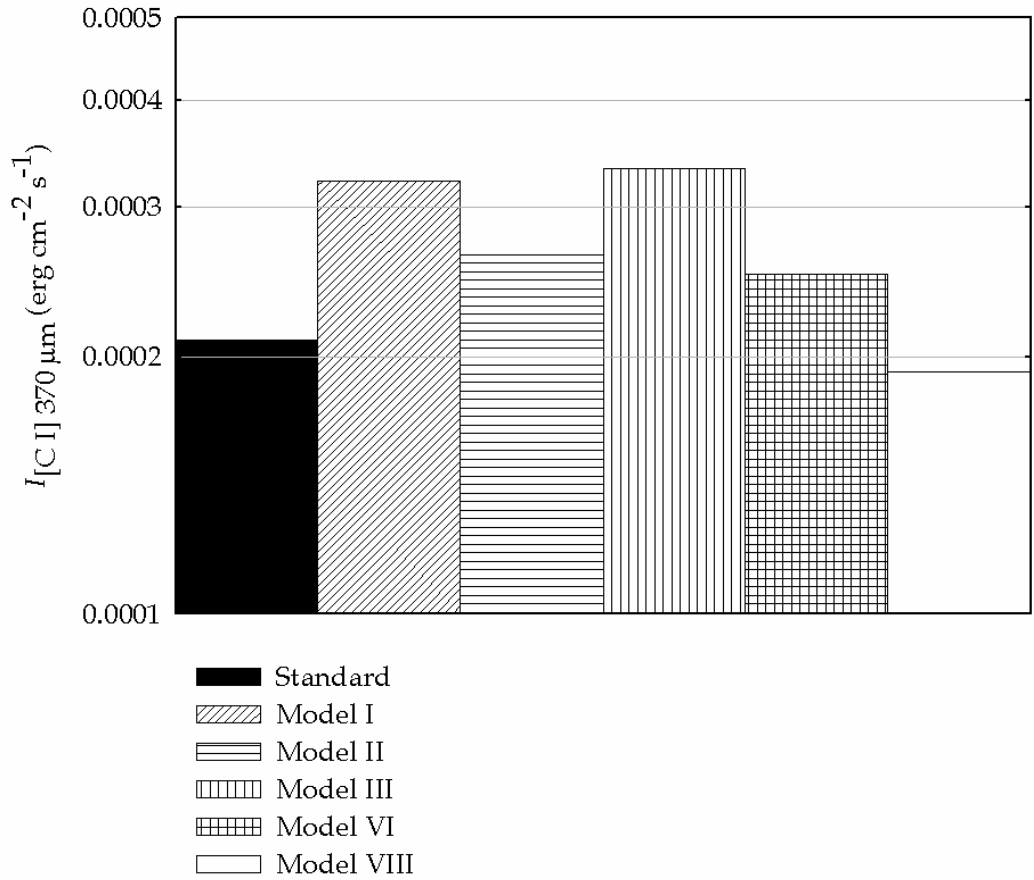

Figure 13 Predicted intensity of the [C I] 370 μm line for various treatments of physical processes in a PDR.

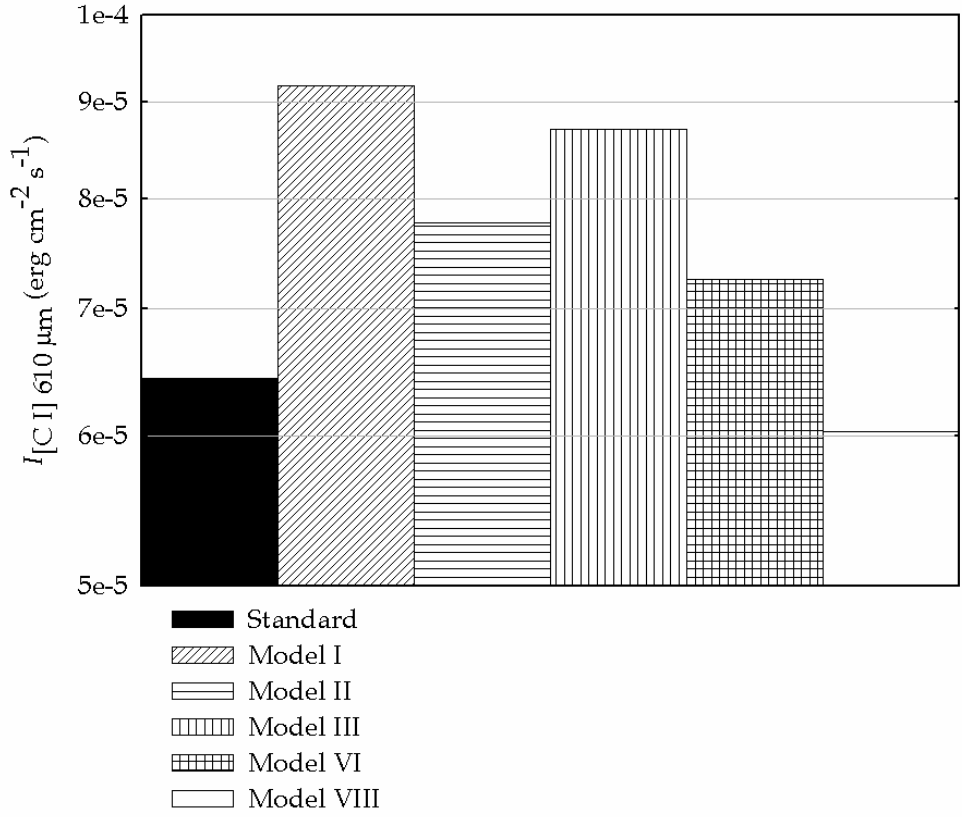

Figure 14 Predicted intensity of the [C I] 610 μm line for various treatments of physical processes in a PDR.

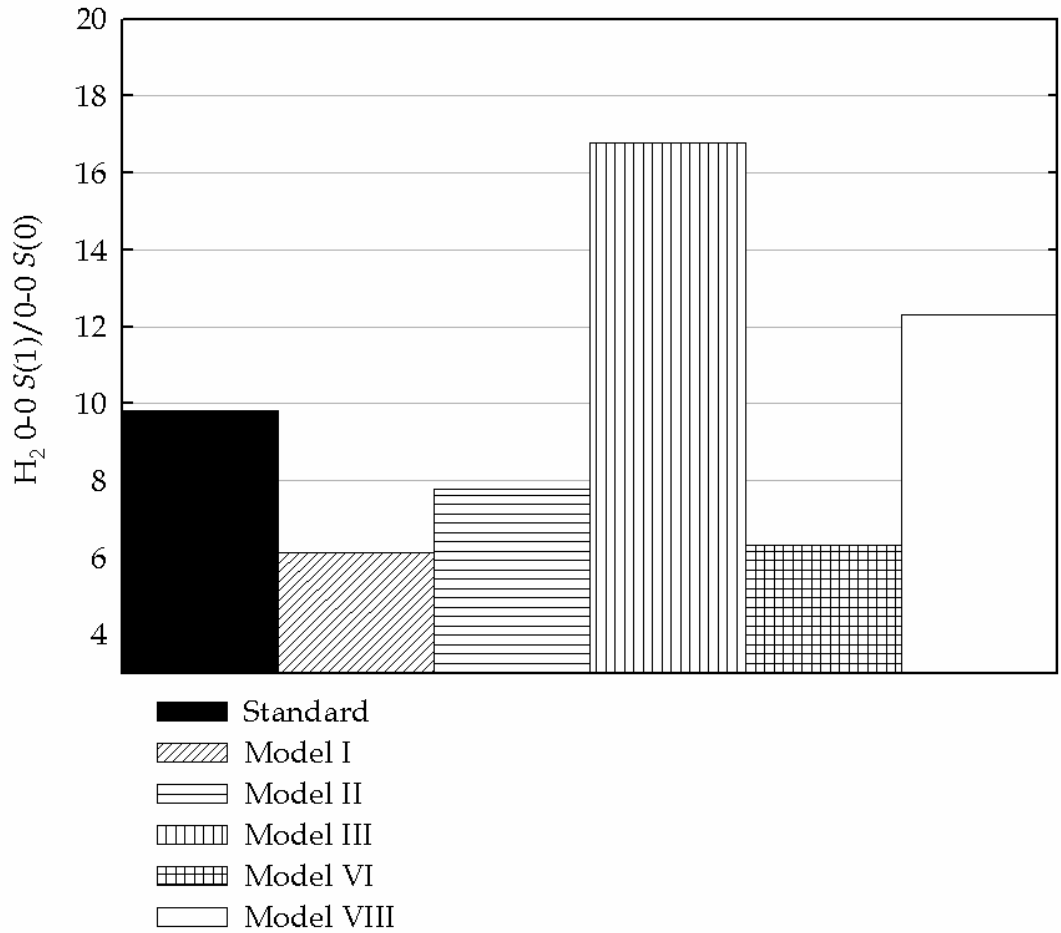

Figure 15 Predicted intensity of the $H_2$ 0-0 S(1)/0-0 S(0) line ratio for various treatments of physical processes in a PDR.

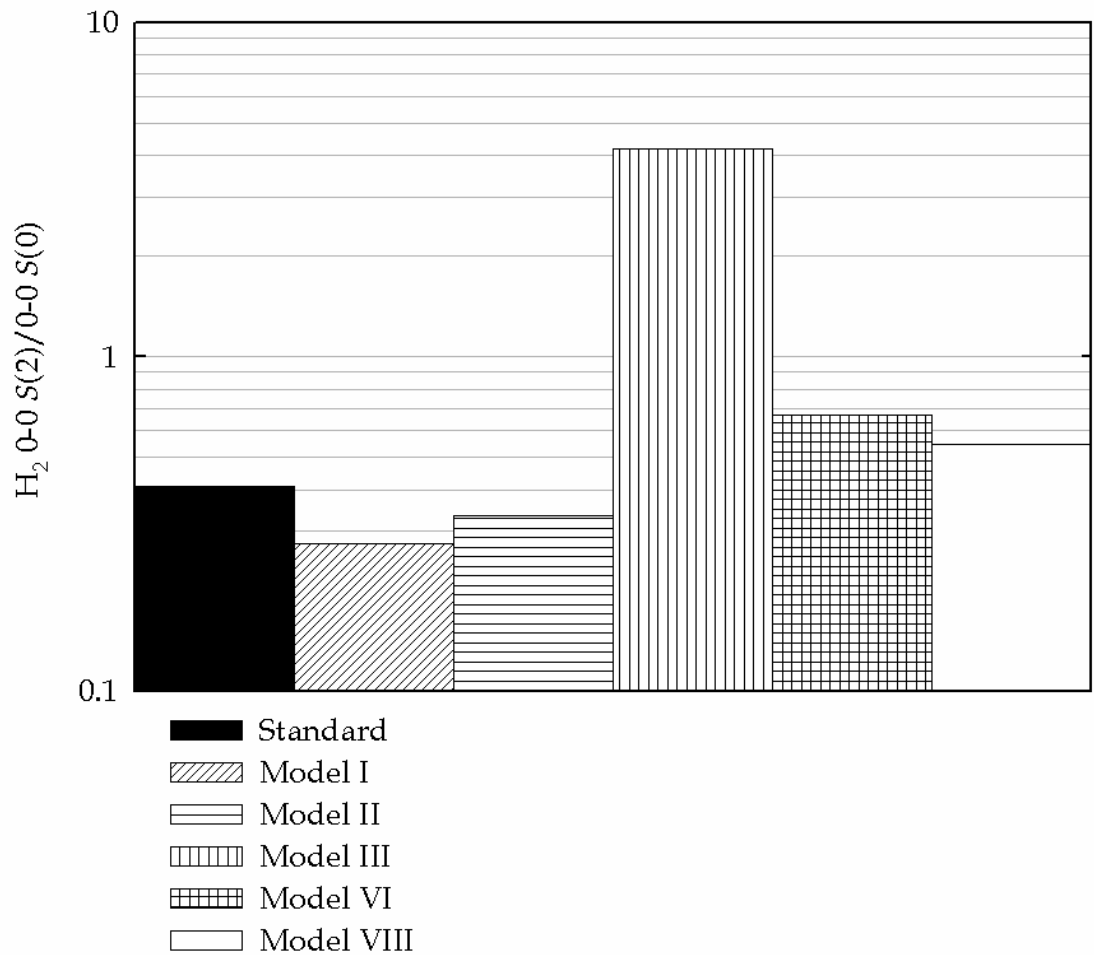

Figure 16 Predicted intensity of the $H_2$ 0-0 S(2)/0-0 S(0) line ratio for various treatments of physical processes in a PDR.

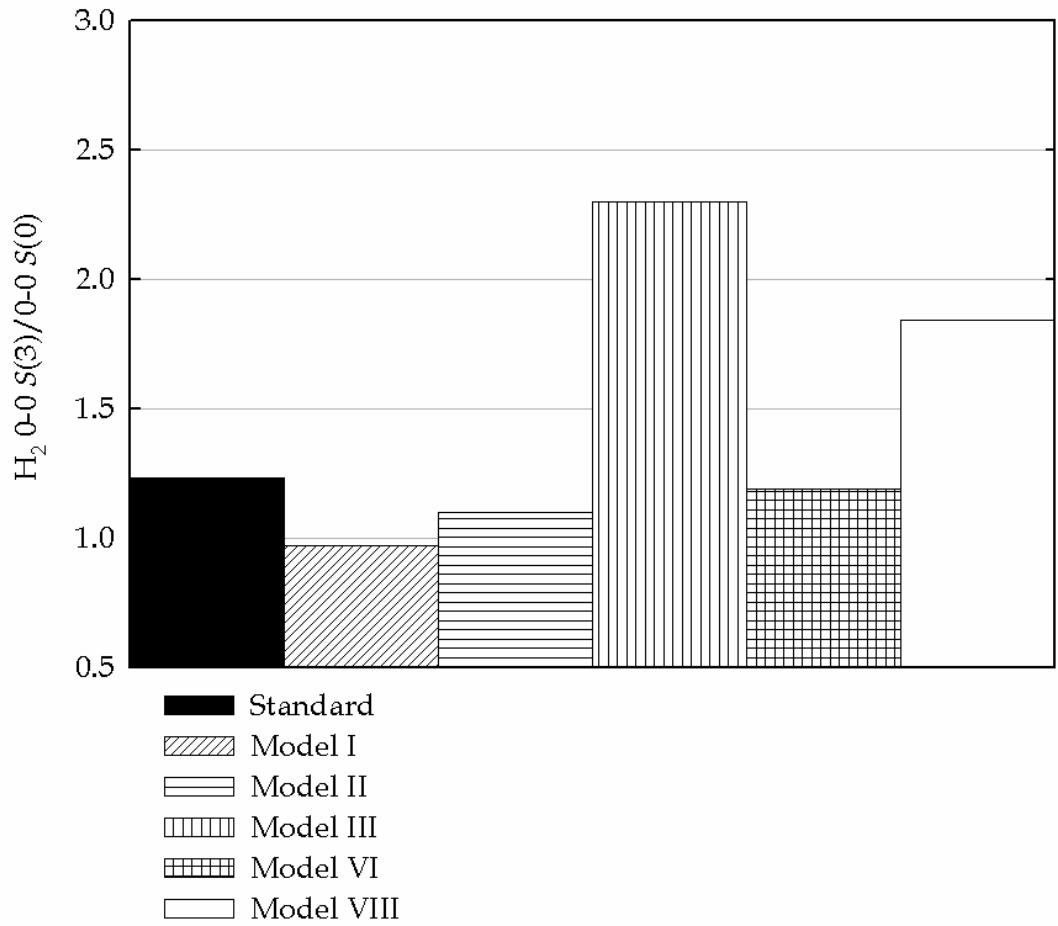

Figure 17 Predicted intensity of the $H_2$ 0-0 S(3)/0-0 S(0) line ratio for various treatments of physical processes in a PDR.